\documentclass[prb, amsmath, amssymb, citeautoscript, reprint]{revtex4-2}
\usepackage{graphicx, float, xcolor, pgf}
\usepackage{physics, siunitx}
\usepackage[caption=false, justification=centerlast]{subfig}

\usepackage{hyperref}
\hypersetup{
    pdfauthor={Amartya Bose},
    pdftitle={Analysis of Time-Evolution of Gaussian Wavepackets in Non-Hermitian Systems},
    colorlinks=true,
}

\allowdisplaybreaks

\begin{document}
\title{Analysis of Time-Evolution of Gaussian Wavepackets in Non-Hermitian Systems}
\author{Amartya Bose}
\affiliation{Department of Chemical Sciences, Tata Institute of Fundamental Research, Mumbai 400005, India}
\email{amartya.bose@tifr.res.in}
\begin{abstract}
    Simulation and analysis of multidimensional dynamics of a quantum
    non-Hmeritian system is a challenging problem. Gaussian wavepacket dynamics
    has proven to be an intuitive semiclassical approach to approximately
    solving the dynamics of quantum systems. A Gaussian wavepacket approach is
    proposed for a continuous space extension to the Hatano-Nelson model that
    enables transparent analysis of the dynamics in terms of complex classical
    trajectories. We demonstrate certain cases where the configuration
    space trajectory can be made fully real by transforming the initial
    conditions to account for the non-Hermiticity appropriately through the
    momentum coordinates. However, in general the complex phase space is
    unavoidable. For the cases where the trajectory is real, the effective force
    can be decomposed into that due to the potential energy surface and that due
    to the imaginary vector potential. The impact of the vector potential on the
    trajectory of the wavepacket is directly proportional to both the strength
    of the vector potential and the width of the wavepacket.
\end{abstract}
\maketitle
\section{Introduction}
Quantum mechanics is built on the foundation of Hermitian operators.
Hermiticity, especially of the Hamiltonian, guarantees a real spectrum,
orthonormal eigenstates, and a unitary time evolution operator. However,
open-quantum systems often incorporate interactions with the environment which
involve gain and loss of energy and / or particles can no longer be modeled
using Hermitian Hamiltonians. Such non-Hermitian systems have been under intense
study over the past few decades~\cite{hatanoVortexPinningNonHermitian1997,
    benderRealSpectraNonHermitian1998, benderGeneralizedPTSymmetry2002,
    benderMakingSenseNonHermitian2007}, with simulations of the dynamics in such
systems becoming a focus in recent
times~\cite{sergiNonHermitianQuantumDynamics2013,
    joshiExceptionalPointsDynamics2018, wangExceptionalPointsDynamics2020,
    liDynamicSkinEffects2022, keFloquetEngineeringNonHermitian2023,
    oritoEntanglementDynamicsManybody2023}.

The wide-ranging applicability and interest of these non-Hermitian systems have
been outlined in various review articles~\cite{zhangReviewNonHermitianSkin2022,
    banerjeeNonHermitianTopologicalPhases2023,
    okumaNonHermitianTopologicalPhenomena2023}. It is notable that despite
Hermiticity being one of the corner-stones of quantum mechanics, over the past
few decades, it has been shown that Hamiltonians that respect the parity-time
($\mathcal{PT}$) symmetry have real
eigenvalues~\cite{benderRealSpectraNonHermitian1998}. This observation has
spawned research into $\mathcal{PT}$-symmetric quantum
mechanics~\cite{benderGeneralizedPTSymmetry2002,
    benderIntroductionSymmetricQuantum2005, benderMakingSenseNonHermitian2007,
    benderPTsymmetricQuantumTheory2015, benderPTSymmetryQuantum2018}. Optics in
particular has proven to be an exceptionally versatile landscape for probing
the effects of $\mathcal{PT}$-symmetry and
non-Hermiticity~\cite{makrisBeamDynamicsMathcal2008,
    musslimaniOpticalSolitonsMathcal2008, ruterObservationParityTime2010,
    regensburgerParityTimeSynthetic2012}. \citet{ruterObservationParityTime2010}
have observed both a spontaneous breaking of the $\mathcal{PT}$-symmetry as well
as a violation of the left-right symmetry in some optical systems.

\citet{liDynamicSkinEffects2022} have recently investigated the time evolution
of a Gaussian wavepacket under a continuum extension of the Hatano-Nelson
model~\cite{hatanoVortexPinningNonHermitian1997} in a finite 1-dimensional box.
In addition to the anomolous reflection from boundaries, that they have dubbed
the ``dynamical'' skin effect, they have also noticed an acceleration being
produced by the non-Hermitian term. These very interesting observations lead
naturally to the question of how the non-Hermiticity in the system ``interacts''
with an external potential. How does the dynamics of a particle in a potential
landscape change due to the non-Hermitian terms. Semiclassical analysis seems to
be invaluable in analysing the dynamics of such complicated systems.

Various methods for simulating the dynamics of a quantum sytem in a
semiclassical manner has been developed over the years. These are generally
obtained by evaluating the Feynman path
integral~\cite{feynmanQuantumMechanicsPath2010} under stationary phase
approximation valid under under the limit of $\hbar\to 0$. This has spawned a
wide range of
approximations~\cite{vanvleckCorrespondencePrincipleStatistical1928,
    wignerQuantumCorrectionThermodynamic1932,
    gutzwillerPhaseIntegralApproximation1967,
    hermanSemiclasicalJustificationUse1984, millerAlternateDerivationHerman2002}.
Gaussian wavepacket dynamics (GWD)~\cite{hellerTimeDependentApproach1975,
    hellerTimeDependentVariational1976,hellerSemiclassicalWayDynamics2018} provides
an alternate, complementary approach to semiclassical simulations where one
tracks the evolution of a Gaussian wavepacket over time and realizes that under
certain conditions the center of the wave packet and the average momentum follow
the classical equations of motion. Recently, GWD, with further approximations,
has been applied to \textit{ab initio}
systems~\cite{begusicOntheflyInitioThree2018,
    begusicSingleHessianThawedGaussian2019, begusicOntheflyInitioSemiclassical2020a,
    begusicApplicabilityThawedGaussian2022}. The original GWD has also been extended
to classical Hamiltonians over a complex phase space to handle classically
forbidden processes~\cite{huberGeneralizedGaussianWave1987,
    huberGeneralizedGaussianWave1988}. This complex version of GWD called the
generalized Gaussian wavepacket dynamics (GGWD) is closely related to path
integrals in coherent state
representation~\cite{weissmanStationaryPhaseEvaluation1983}.

In this work we present the thawed Gaussian dynamics corresponding to a
generalized Hatano-Nelson (HN) continuum model under an external potential. The
non-Hermitian term appears as an imaginary vector potential. Despite the initial
average position and momentum corresponding to a Gaussian wavepacket being real,
the imaginary vector potential ensures that the classical trajectory goes over a
complex phase space. Thus, in general, the method requires complex phase space
classical trajectories and corresponding generalized Gaussians. However, we show
that it is possible to use the properties of these general Gaussians to
transform the initial momentum into the complex plane dictated by the imaginary
vector potential, which ensures that the initial position and velocity are real.
Consequently, the corresponding configuration space trajectory, which is called
the ``guiding'' trajectory, becomes completely real and identical to the
classical trajectory in absence of the non-Hermitian term. The real phase space
center of the wavepacket is obtained as a transformation of this real guiding
trajectory. We also show that the deviation of the real center of the wavepacket
from the guiding trajectory is related to the width of the wavepacket, which in
turn depends on the non-Hermitian terms.

While the results present here are exact for quadratic external potentials and
linear imaginary vector potential, higher order potentials are dealt with at a
local harmonic level. The paper is organized as follows: Sec.~\ref{sec:method}
outlines the model and develops the non-Hermitian thawed Gaussian wavepacket
dynamics (NH-GWD) approach. Numerical demonstrations of the method are provided
in Sec.~\ref{sec:results}. Finally, some concluding remarks and future
directions are provided in Sec.~\ref{sec:conclusion}.

\section{Model \& Non-Hermitian Thawed Gaussian Wavepacket Dynamics}\label{sec:method}
Consider a generalized HN-inspired continuum model given
by~\cite{hatanoVortexPinningNonHermitian1997}:
\begin{align}
    H & = \sum_j \frac{1}{2m_j}\left(P_j + i b_j(X_j)\right)^2 + V(\vb{X})                                                                                              \\
      & = \sum_j \left(-\frac{\hbar^2}{2m_j}\partial_j^2 + \frac{\hbar b_j(x_j)}{m_j}\partial_j + \frac{\hbar}{2m_j}b'_j(x_j) - \frac{b_j^2(x_j)}{2m_j}\right)\nonumber \\
      & + V(\vb{x})\label{eq:pos_space_NH}
\end{align}
where $X_j, P_j$ are the position and momentum operators along the $j$th
dimension, $b_j$ is a real function that brings in the non-Hermiticity, and
$\partial_j = \pdv{x_j}$. The non-Hermitian part, brought in by the $b$
functions, can be interpreted as an imaginary vector potential. Discretizing
Eq.~\ref{eq:pos_space_NH} on a lattice reveals that $b_j$ causes the
left-hopping amplitude and the right-hopping amplitude along the $j$th dimension
to be different making the system non-Hermitian.

We want to understand the dynamics of a generalized Gaussian wavepacket. We
assume that the propagated wavepacket retains its Gaussian form given
by~\cite{huberGeneralizedGaussianWave1988}
\begin{align}
    \braket{\vb{x}}{\alpha, \vb{q}, \vb{p}, \gamma, t} & = \exp\left(\frac{i}{\hbar}\left(\sum_{ij}\alpha_{ij}(t) (x_i - q_i(t)) (x_j - q_j(t))\right.\right. \nonumber \\
                                                       & \left.\left. + \sum_j p_j(t) (x_j - q_j(t)) + \gamma(t)\right)\right)
\end{align}
where the time-dependent parameters, $\alpha_{ij}(t)$, $q_j(t)$, $p_j(t)$, and
$\gamma(t)$ are in general all complex. The imaginary part of $\alpha_{ij}$ is
related to the width of the wavepacket ($\sigma =
    \sqrt{\frac{\hbar}{2\Im\alpha}}$). If $q_j(t)$ and $p_j(t)$ are real, they
represent the centre of the wavepacket and the momentum respectively. The term
$\gamma(t)$ represents the time-dependent phase of the wave packet. Every
generalized Gaussian wavepackets does not uniquely represent one wave function.
Two wavepackets are the same if the following two quantities are the same for
both of them:
\begin{align}
    -2\sum_{j}\alpha_{ij}(t)q_j(t) + p_i(t)                                & = \text{const},\label{eq:pos_mom_invariance} \\
    \gamma(t) + \sum_{ij}\alpha_{ij}(t)q_i(t)q_j(t) - \sum_j p_j(t) q_j(t) & = \text{const}.\label{eq:gamma_invariance}
\end{align}

The position and momentum parameters in a generalized Gaussian wavepacket do not
correspond directly to the physically relevant phase space location of the
wavepacket, which are real quantities. By considering the real and imaginary
parts of Eq.~\ref{eq:pos_mom_invariance}, conversion from the complex phase
space to the real phase space is enabled
by~\cite{huberGeneralizedGaussianWave1987}
\begin{align}
    \vb{Q}(t) & = \Re\vb{q}(t) + (\Im\alpha(t))^{-1}\left(\Re\alpha(t)\Im\vb{q}(t) - \frac{1}{2}\Im\vb{p}(t)\right)\label{eq:real_pos} \\
    \vb{P}(t) & = \Re\vb{p}(t) + 2\Im\alpha(t)\Im\vb{q}(t) + 2\Re\alpha(t)\left(\vb{Q}(t) - \vb{q}(t)\right)\label{eq:real_mom}
\end{align}

There are several restrictions to GWD: first, it is exact for upto quadratic
potentials; second, it cannot in general account for classically forbidden
processes; and third, if the initial state is a superposition of Gaussians, the
norm is not necessarily conserved. The non-uniqueness of the generalized Gaussian
wavepackets can be leveraged to overcome these restrictions by choosing a
complex phase space initial condition taylored to the final
state~\cite{huberGeneralizedGaussianWave1987, huberGeneralizedGaussianWave1988}.

GWD assumes that the potential can treated under a local harmonic approximation:
\begin{align}
    V(\vb{x}) & = V(\vb{q}) + V^{(1)}(\vb{q}) \cdot (\vb{x} - \vb{q})\nonumber    \\
              & + (\vb{x} - \vb{q}) \cdot V^{(2)}(\vb{q}) \cdot (\vb{x} - \vb{q})
\end{align}
which works if the potential is relatively slowly varying over the width of the
wavepacket. Because $b_j^2$ contributes to an effective potential, for this work
we assume a locally linear imaginary potential,
\begin{align}
    b_j(x_j) & = b_j(q_j) + b_j^{(1)}(q_j) (x_j - q_j),
\end{align}
which makes $b_j^2$ a locally quadratic function.

Solving the time-dependent Schr\"odinger equation,
\begin{align}
    i\hbar\dv{\ket{\alpha, \vb{q}, \vb{p}, \gamma, t}}{t} & = H\ket{\alpha, \vb{q}, \vb{p}, \gamma, t},
\end{align}
we get equations of motion (EOM) for $\alpha$, $\vb{q}$, and $\vb{p}$
\begin{widetext}
    \begin{align}
        \sum_l 2\alpha_{jl}\dot{q}_l(t) - \dot{p}_j(t) & = V^{(1)}_j(\vb{q}(t)) - \frac{b_j(q_j(t))b_j^{(1)}(q_j(t))}{m_j} + \sum_k\frac{2\alpha_{jk}(t)}{m_k}(p_k(t) + ib_k(q_k(t))) + \frac{ip_j(t)b_j^{(1)}(q_j)}{m_j}\label{eq:eom1}                        \\
        -\dot{\alpha}_{kl}                             & = V^{(2)}_{kl}(\vb{q}(t)) - \frac{b_k^{(1)2}(q(t))}{2m_k}\delta_{kl} + 2i\frac{\alpha_{kl}(t)b_l^{(1)}(q_l)}{m_l} + 2\sum_j \frac{\alpha_{kj}(t)\alpha_{jl}(t)}{m_j}\label{eq:eom2}                    \\
        \sum_j p_j(t)\dot{q}_j(t) - \dot{\gamma}(t)    & = V(\vb{q}(t)) + \sum_j \left(\hbar\frac{b^{(1)}_j(q_j(t))}{m_j} - \frac{b_j^{2}(q_j(t))}{2m_j} + \frac{ip_j(t)b_j(q_j(t))}{m_j} + \frac{p_j^2(t) - 2i\hbar\alpha_{jj}(t)}{2m_j}\right)\label{eq:eom3}
    \end{align}
\end{widetext}

Consider the classical equations of motion under the Hamiltonian Eq.~\ref{eq:pos_space_NH},
\begin{align}
    \dot{q}_j(t) & = \frac{p_j + ib_j(q_j)}{m_j}                                                      \label{eq:hamil_pos} \\
    \dot{p}_j(t) & = -\partial_j V(\vb{q}) - \frac{ib_j^{(1)}(q_j)}{m_j}\left(p_j + ib_j(q_j)\right).\label{eq:hamil_mom}
\end{align}
These classical equations satisfy Eq.~\ref{eq:eom1}. However, notice that the
imaginary vector potential makes the classical trajectory, $(\vb{q}(t),
    \vb{p}(t))$, are in general complex, even if the initial conditions,
$(\vb{q}(0), \vb{p}(0))$, are real. Using Eq.~\ref{eq:hamil_pos}, we can
simplify Eq.~\ref{eq:eom3} to
\begin{align}
    \dot{\gamma}(t) & = \sum_j \frac{p_j^2(t)}{2m_j} - V(\vb{q}(t)) + i\hbar\sum_j\frac{\alpha_{jj}(t)}{m_j}\nonumber \\
                    & + \sum_j \frac{b_j^2(q_j(t))-2\hbar b^{(1)}_j(q_j(t))}{2m_j}\label{eq:eom3_simple}
\end{align}
So we now have the full EOM for the Gaussian wavepacket
given by Eqs.~\ref{eq:eom2}, and~\ref{eq:hamil_pos} -- \ref{eq:eom3_simple}.
Notice that from Eqs.~\ref{eq:hamil_pos} and~\ref{eq:hamil_mom}, though the
initial position and momenta start off being real, the equations of motion make
them complex. We will now show that for the cases of a constant and a linear
imaginary vector potential, it is possible to transform the initial conditions
in such a manner that the trajectory becomes real in configuration space. This
``guiding'' trajectory will allow us to decompose the effects of the external
potential from that of the non-Hermitian terms in an intuitive manner.

\subsection{Constant imaginary vector potential}
Suppose $b_j(q_j) = k_j$ is a constant. This is the case that has been recently
studied by Li and Wan~\cite{liDynamicSkinEffects2022} with $V=0$. For the
generalized Gaussian wavepacket, the EOM become:
\begin{align}
    \dot{q}_j(t)       & = \frac{p_j + i k_j}{m_j}\label{eq:vel_const_im_pot}                                                               \\
    \dot{p}_j(t)       & = -\partial_j V(\vb{q})                                                                                            \\
    -\dot{\alpha}_{kl} & = V^{(2)}_{kl}(\vb{q}(t)) + 2\sum_j \frac{\alpha_{kj}(t)\alpha_{jl}(t)}{m_j}                                       \\
    \dot{\gamma}(t)    & = \sum_j \frac{p_j^2(t)}{2m_j} - V(\vb{q}(t)) + i\hbar\sum_j\frac{\alpha_{jj}(t)}{m_j} + \sum_j \frac{k_j^2}{2m_j}
\end{align}
Generally, $\vb{q}(0)$ and $\vb{p}(0)$ are both real. Consequently using
Eq.~\ref{eq:vel_const_im_pot}, we find that the initial velocity is complex.
This of course leads to classical trajectories that are complex in both position
and momentum dimensions. This problem can solved by analytically continuing the
potential to a complex plane~\cite{huberGeneralizedGaussianWave1987}. For the
typical wavepacket, the initial value of $\alpha$ is usually imaginary
($\alpha(0) = \frac{i\hbar}{2\sigma^2}$).

However, let us see if we can construct real trajectories in some way. Notice at
this stage that, if we arbitrarily choose a new, in general complex, initial
momentum $\vb{\tilde{p}}(0)$, the physics would remain the same if, using
Eq.~\ref{eq:pos_mom_invariance}, a corresponding new position is chosen as
follows:
\begin{align}
    \vb{\tilde{q}}(0) & = \vb{q}(0) + \frac{1}{2}\alpha(0)^{-1}\left(\vb{\tilde{p}}(0) - \vb{p}(0)\right),
\end{align}
and $\gamma(0)$ is changed according to Eq.~\ref{eq:gamma_invariance}. If
$\vb{q}(0)$ is real, $\vb{\tilde{q}}(0)$ is real if $\vb{\tilde{p}}(0) -
    \vb{p}(0)$ is imaginary because $\alpha(0)$ is initially imaginary.
Additionally,
\begin{align}
    \vb{\dot{\tilde{q}}}(0) & = m^{-1} \left(\vb{\tilde{p}}(0) + i\vb{k}\right).
\end{align}
Now, if we set $\vb{\tilde{p}}(0) = \vb{p}(0) - i\vb{k}$, then both
$\vb{\tilde{q}}(0) = \vb{q}(0) - \frac{i}{2} (\alpha(0))^{-1} \vb{k}$ and
$\vb{\dot{\tilde{q}}}(0) = m^{-1}\vb{p}(0)$ both become real. With such a
transformation of the initial positions and momenta into a generalized Gaussian
wavepacket, the usual classical equations of motion,
\begin{align}
    m \vb{\ddot{\tilde{q}}} & = -\partial_j V(\vb{\tilde{q}}),     \label{eq:class_eom} \\
    \vb{\tilde{p}}(t)       & = m\vb{\dot{\tilde{q}}}(t) - i\vb{k}
\end{align}
give real positions and velocities with no impact from the imaginary vector
potential $i\vb{k}$. This trajectory, $\vb{\tilde{q}}(t)$ which is uniquely
determined only via the external potential, is the ``guiding'' classical
trajectory. However, the corresponding momenta are complex. Converting back to
real phase space coordinates using Eqs.~\ref{eq:real_pos} and~\ref{eq:real_mom},
one obtains,
\begin{align}
    \vb{Q}(t) & = \vb{\tilde{q}}(t) + \frac{1}{2} (\Im\alpha(t))^{-1} \vb{k}\label{eq:real_trajectory_pos}        \\
    \vb{P}(t) & = m\vb{\dot{\tilde{q}}}(t) + \Re\alpha(t) (\Im\alpha(t))^{-1}\vb{k}\label{eq:real_trajectory_mom}
\end{align}
With this, we have solved the problem for a constant imaginary vector potential.
This solution is exact for systems upto a quadratic potential. Beyond that this
effectively treats the potential under a local harmonic approximation. While the
full generalized Gaussian wavepacket analysis of these systems following Huber
and Heller~\cite{huberGeneralizedGaussianWave1987} will be done in a future
work, the current development already gives us different perspectives to analyse
the problem from.

Notice that the classical equations of motion, Eq.~\ref{eq:class_eom} for the
guiding trajectory, knows nothing about $\vb{k}$, it follows the forces from the
potential energy landscape.The equation of motion of $\alpha(t)$,
Eq.~\ref{eq:free_alpha}, is independent of $\vb{k}$, which in this case is zero.
So, $\alpha(t)$ only depends on the external potential and the guiding
trajectory. It also knows nothing about the non-Hermitian term. The imaginary vector
potential enters the trajectory in Eq.~\ref{eq:real_trajectory_pos}, when we
insist upon a fully real phasespace. The prefactor of $(\Im\alpha(t))^{-1}$ is
proportional to the dynamical width of the wavepacket $\sigma(t)$. Thus the
impact of $\vb{k}$ increases as the wavepacket gets broader.

Now, let us obtain the analytical solution of some known cases.

\subsubsection{1D Free Particle}
Let us obtain the analytic results for some of the simple cases. First consider a
free particle in one dimension, $V(q) = 0$ as explored in
Ref.~\cite{liDynamicSkinEffects2022}, the solution
is~\cite{hellerTimeDependentApproach1975}:
\begin{align}
    \tilde{q}(t) & = \tilde{q}(0) + p(0)t / m\label{eq:free_traj}                                                                                       \\
    \alpha(t)    & =\frac{m\alpha(0)}{2\alpha(0)t + m} = \frac{im\hbar}{2i\hbar t + 2m\sigma^2(0)},                                                     \\
                 & = \frac{m\hbar^2 t}{2m^2\sigma^4(0) + 2\hbar^2t^2} + \frac{im^2\hbar\sigma^2(0)}{2m^2\sigma^4(0) + 2\hbar^2t^2}\label{eq:free_alpha}
\end{align}
where $\sigma(0)$ is the initial standard deviation of the wavepacket.
Consequently,
\begin{align}
    Q(t) & = q(0) + \frac{p(0)}{m}t + \frac{\hbar k}{m^2\sigma^2(0)}t^2 \\
    P(t) & = p(0) + \frac{\hbar k}{m\sigma^2(0)} t
\end{align}
Notice that the only acceleration term comes because of the imaginary potential
$k$ that acts like a constant external force.

So, it is natural to be curious if this force can be ``counteracted,'' if we
have a linear external potential leading to a constant force.

\subsubsection{1D Linear Ramp}
Consider the case where $V(x) = \beta x$. For this case, we will have:
\begin{align}
    p(t)         & = p(0) - \beta t                                         \\
    \tilde{q}(t) & = \tilde{q}(0) + \frac{p(0)}{m} t - \frac{\beta}{2m}t^2.
\end{align}
Since the potential is linear, the solution for $\alpha(t)$ remains the same as
in Eq.~\ref{eq:free_alpha}. Therefore, the real phase space location of the wave
packet will be given by:
\begin{align}
    Q(t) & =\tilde{q}(t) + \left(\frac{\sigma^2(0)}{\hbar} + \frac{\hbar t^2}{m^2\sigma^2(0)}\right)k    \\
         & = q(0) + \frac{p(0)}{m}t + \left(\frac{\hbar k}{m^2\sigma^2(0)} - \frac{\beta}{2m}\right)t^2.
\end{align}
Therefore, if we choose $\beta = \frac{2\hbar k}{m\sigma^2(0)}$, the wavepackets
center would behave as a free particle. We need to be careful at this stage. The
``additive'' nature of the forces due to the constant imaginary vector potential
and the external potential field holds in this case because $\alpha(t)$ is not
affected by the potential. For quadratic and higher potentials where $\alpha(t)$
itself would change, this feature will not appear.

\subsubsection{1D Harmonic Oscillator}
Next, we obtain the solutions for a harmonic $V(x) = \frac{1}{2}m\omega^2 x^2$.
It is well known that GWD is exact for these cases with the solution being given
by~\cite{hellerTimeDependentApproach1975}:
\begin{align}
    \tilde{q}(t) & = \tilde{q}(0)\cos(\omega t) + \frac{p(0)}{m\omega}\sin(\omega t)                                                                                  \\
    \alpha(t)    & = -\frac{m\omega}{2}\left(\frac{\frac{1}{2}m\omega - \alpha(0) \cot(\omega t)}{\alpha(0) + \frac{1}{2}m\omega\cot(\omega t)}\right)                \\
                 & = -\frac{m\omega}{2}\left(\frac{m^2\omega^2\sigma^4(0) - \hbar^2}{\hbar^2\tan^2(\omega t) + m^2\omega^2\sigma^4(0)\cot(\omega t)}\right. \nonumber \\
                 & \left. -i\frac{\hbar m\omega\sigma^2(0)}{\hbar^2\sin^2(\omega t) + m^2\omega^2\sigma^4(0)\cos^2(\omega t)}\right)
\end{align}
Therefore, the center of the wavepacket goes as
\begin{align}
    Q(t) & = \tilde{q}(t) + \left(\frac{\hbar^2\sin^2(\omega t) + m^2\omega^2\sigma^4(0)\cos^2(\omega t)}{\hbar m^2 \omega^2\sigma^2(0)}\right) k.
\end{align}
Notice that $\tilde{q}(t)$ is the guiding classical trajectory that already has
the free harmonic oscillations with frequency $\omega$. The imaginary vector
potential brings in an extra component of frequency $2\omega$ in the form of
terms with $\sin^2(\omega t)$ and $\cos^2(\omega t)$.

\subsection{Linear imaginary vector potential}
We have shown that for a constant imaginary vector potential, it is possible to
transform the initial condition to obtain trajectories that a completely real in
the configuration space. Can a similar transformation be done for the case when
$\vb{b}(\vb{q}) = k\vb{q} + \vb{c}$?

Consider a wavepacket with initial phase space positions $\vb{q}(0)$ and
$\vb{p}(0)$. While these are real, the problem is once again, that the
trajectory would be come complex because of the imaginary vector potential,
which renders $\dot{\vb{q}}(0)$ complex. Let us suppose that $\tilde{\vb{p}}(0)$
is the momentum that allows both the position and velocity to be real. Let the
corresponding position be $\tilde{\vb{q}}(0)$. As in the previous section,
\begin{align}
    \tilde{\vb{q}}(0)       & = \vb{q}(0) + \frac{1}{2}\alpha(0)^{-1}\left(\tilde{\vb{p}}(0) - \vb{p}(0)\right) \\
    \vb{\dot{\tilde{q}}}(0) & = m^{-1} \left(\vb{\tilde{p}}(0) + i\vb{b}(\vb{\tilde{q}}(0))\right)
\end{align}
For $\tilde{\vb{q}}(0)$ to be real, $\vb{\tilde{p}}(0) - \vb{p}(0)$ has to be
imaginary. Let $\vb{\tilde{p}}(0) = \vb{p}(0) + i\vb{f}$, where $i\vb{f} =
    2\alpha(0)(\vb{\tilde{q}}(0) - \vb{q}(0))$. This leads to

\begin{align}
    \vb{\dot{\tilde{q}}}(0) & = m^{-1}\left(\vb{p}(0) + 2\alpha(0)(\vb{\tilde{q}}(0) - \vb{q}(0)) + i\vb{b}(\vb{\tilde{q}}(0))\right)
\end{align}
Given that $\alpha(0)$ is imaginary, $\vb{\dot{\tilde{q}}}(0)$ can be real only
if the sum of the second and third terms is zero. This condition is equivalent
to the following:
\begin{align}
    \vb{\tilde{q}}(0) & = (2\Im\alpha(0) + k)^{-1}(2\Im\alpha(0)\vb{q}(0) - \vb{c}) \\
    i\vb{f}           & = -2\alpha(0)(2\Im\alpha(0) + k)^{-1}\vb{b}(\vb{q}(0))
\end{align}

With these initial conditions, the guiding classical trajectory has all real
positions. Once again the imaginary part of the canonical is given by the
imaginary vector potential evaluated at the guiding trajectory. The real phase
space center of the wavepacket can be once again evaluated by
Eqs.~\ref{eq:real_pos} and~\ref{eq:real_mom}. The EOMs for $\alpha(t)$ and
$\gamma(t)$ involve $\vb{b}(\vb{q})$ explicitly.
\begin{widetext}
    \begin{minipage}{\textwidth}
        \begin{figure}[H]
            \vspace{-1.101cm}
            \centering
            \subfloat[$q(0)=0.0, p(0)=0.0$.]{\includegraphics{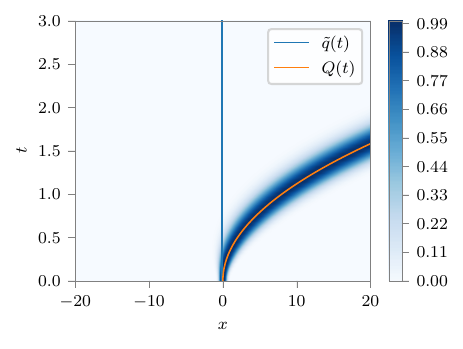}}
            ~\subfloat[$q(0)=0.0, p(0)=-10.0$.]{\includegraphics{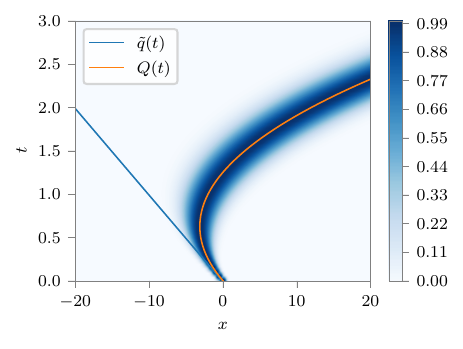}}

            \subfloat[$q(0)=0.0, p(0)=0.0$.]{\includegraphics{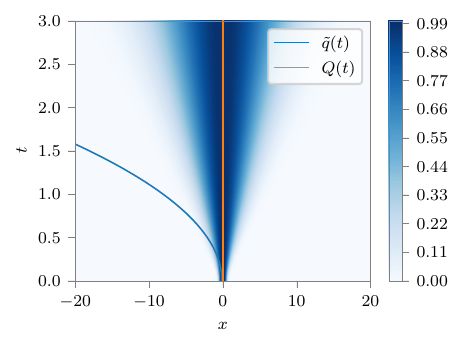}}
            ~\subfloat[$q(0)=0.0, p(0)=-10.0$.]{\includegraphics{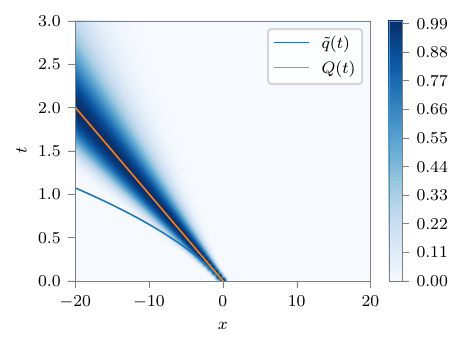}}
            \caption{Plot of $\absolutevalue{\psi(t, x)}^2$ as a function of position and time normalized to a maximum value of 1. The value of the imaginary potential $k=1$ and $m=1$. Top row: free particle; Bottom row: linear ramp with $\beta = \frac{2\hbar k}{m\sigma^2(0)}$.}\label{fig:free_particle_linear_ramp}
        \end{figure}
    \end{minipage}
\end{widetext}
\section{Results}\label{sec:results}
First, we consider the case of a constant imaginary vector potential. In all
these cases, we consider a wavepacket with $\alpha(0) = 4i$. In
Fig.~\ref{fig:free_particle_linear_ramp}, we show the dynamics of a wavepacket
localized originally at $q(0) = 0.0$ with different initial momenta under the
non-Hermitian term, $k$, and a variable external potential. For the case of the
free particle, Fig.~\ref{fig:free_particle_linear_ramp}~(a) and (b), our results
match the analytic solutions in Ref.~\cite{liDynamicSkinEffects2022}. For the
linear ramp, we have picked $\beta$ to be at the critical value of $\frac{2\hbar
        k}{m\sigma^2(0)}$. As a consequence, notice that in
Fig.~\ref{fig:free_particle_linear_ramp}~(c) and (d) we only see the center,
$Q(t)$, moving ``unforced'' in a straight line.

\begin{figure}
    \centering
    \subfloat[$q(0)=0.0, p(0)=0.0$.]{\includegraphics{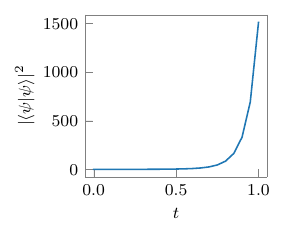}}
    ~\subfloat[$q(0)=0.0, p(0)=-10.0$.]{\includegraphics{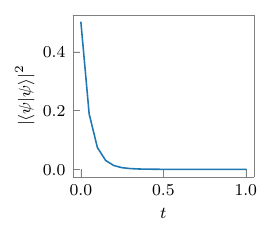}}

    \subfloat[$q(0)=0.0, p(0)=0.0$.]{\includegraphics{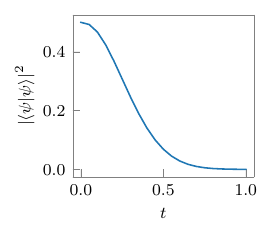}}
    ~\subfloat[$q(0)=0.0, p(0)=-10.0$.]{\includegraphics{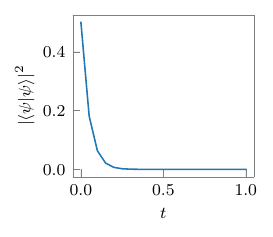}}
    \caption{Plot of $\absolutevalue{\braket{\psi}}^2$. Subfigures correspond to Fig.~\ref{fig:free_particle_linear_ramp}.}\label{fig:free_particle_linear_ramp_norm}
\end{figure}

The magnitude of the wave function corresponding to the particle with no
external potential and the linear ramp are shown in
Fig.~\ref{fig:free_particle_linear_ramp_norm}. While for the free particle, the
norm increases, for the system in an external linear potential, the norm drops
to zero. As is observed, the evolution of the magnitude of the wavefunction
depends upon an interaction of the external potential and the imaginary vector
potential.

\begin{figure}
    \centering
    \subfloat[$q(0)=0.0, p(0)=0.0$.]{\includegraphics{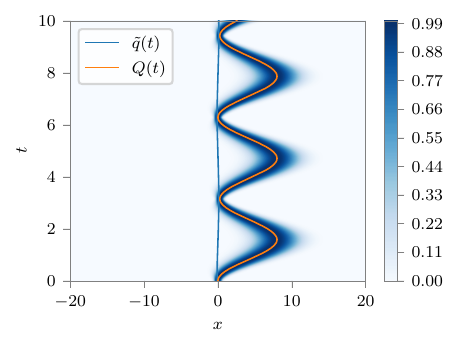}}

    \subfloat[$q(0)=0.0, p(0)=-10.0$.]{\includegraphics{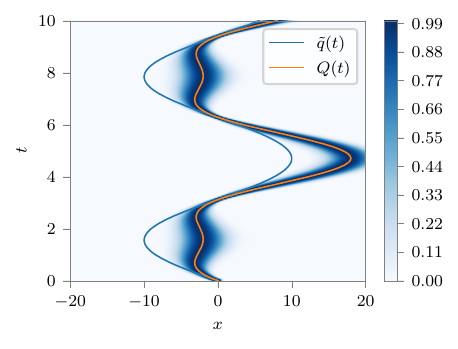}}
    \caption{Plot of $\absolutevalue{\psi(t, x)}^2$ as a function of position and time normalized to a maximum value of 1 for a wave packet moving under an external harmonic potential with $\omega = 1$.}\label{fig:harmonic_GWD}
\end{figure}

\begin{figure}
    \centering
    \subfloat[$q(0)=0.0, p(0)=0.0$.]{\includegraphics{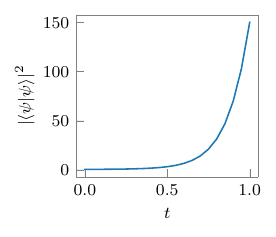}}
    ~\subfloat[$q(0)=0.0, p(0)=-10.0$.]{\includegraphics{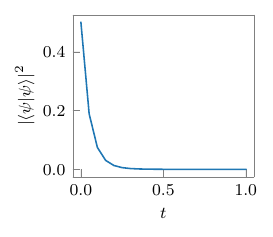}}
    \caption{Plot of norm of the wavepacket as it propagates through the external harmonic potential.}\label{fig:harmonic_norm}
\end{figure}

The dynamics gets interesting when we consider the particle in an external
harmonic potential. The wavepacket is shown in Fig.~\ref{fig:harmonic_GWD}.
Notice that while in the case where the initial condition has 0 momentum and
starts at the origin (Fig.~\ref{fig:harmonic_GWD}~(a)), the dynamics has a
single frequency, the one with a negative momentum
(Fig.~\ref{fig:harmonic_GWD}~(b)) shows an extra ``bow''-like feature that is
phase shifted with respect to the main oscillations. As a result of an
interaction between the constant vector potential and the external harmonic
potential, the dynamics seems to change between the two cases with different
momenta.

The magnitude of the wavepacket for these two cases is shown in
Fig.~\ref{fig:harmonic_norm}. Notice the periodicity of the norm for the case
where the particle was initially stationary. In comparison, no such periodicity
is seen in the case of the particle with $p(0) = -10.0$ within the time period
of $t=10$.

\begin{figure}
    \centering
    \subfloat[$q(0)=0.0, p(0)=0.0$.]{\includegraphics{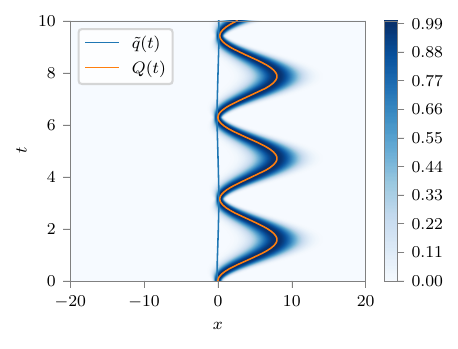}}

    \subfloat[$q(0)=0.0, p(0)=-10.0$.]{\includegraphics{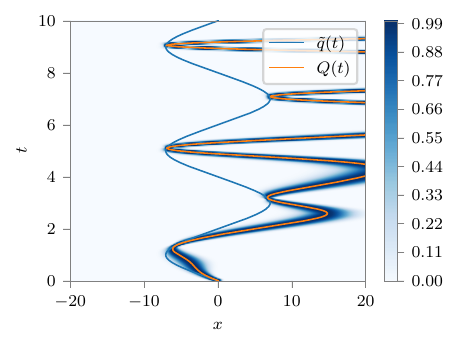}}
    \caption{Plot of $\absolutevalue{\psi(t, x)}^2$ as a function of position and time normalized to a maximum value of 1 for a wave packet moving under an external potential with $V(x) = \frac{1}{2}x^2 + 0.01x^4$.}\label{fig:anharmonic_GWD}
\end{figure}

Till now we have considered the dynamics under potentials that are quadratic or
lower in power. In these cases, GWD is known to be exact. Let us consider $V(x)
    = \frac{1}{2}x^2 + 0.01x^4$. The dynamics for this anharmonic potential is
shown for the same initial conditions in Fig.~\ref{fig:anharmonic_GWD}
simulated using a local harmonic approximation. As the anharmonicity gets
stronger the dynamics would get increasingly approximate. One can
significantly improve the quality of the simulation by using the generalized
Gaussian wavepacket method.

\begin{figure}
    \centering
    \subfloat[$q(0)=0.0, p(0)=0.0$.]{\includegraphics{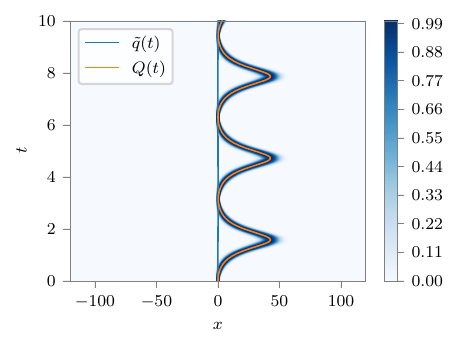}}

    \subfloat[$q(0)=0.0, p(0)=-10.0$.]{\includegraphics{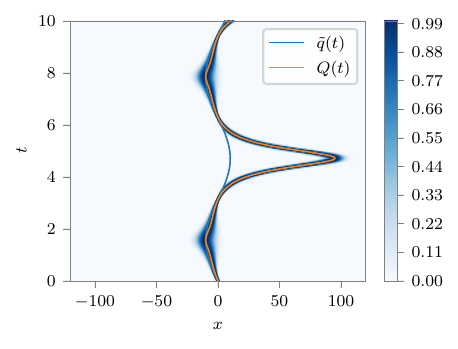}}
    \caption{Plot of $\absolutevalue{\psi(t, x)}^2$ as a function of position and time normalized to a maximum value of 1 for a wave packet moving under an external harmonic potential with $\omega = 1$. The imaginary vector potential is $b(x) = 0.1x + 1.0$.}\label{fig:harmonic_lin_GWD}
\end{figure}
Next, consider a case with a linear imaginary vector potential $b(x) =
    0.1x + 1.0$. We want to understand the interaction of this linear $b$ with an
external harmonic potential $V(x) = \frac{1}{2}x^2$.
Figure~\ref{fig:harmonic_lin_GWD} shows the evolution of the wavepacket in this
case. Notice the profound differences brought in by the change of the imaginary
vector potential.

Finally, recent work has shown the existence of dynamical skin
effect~\cite{liDynamicSkinEffects2022}, where the reflection of a wavepacket
from a hard boundary does not conserve the momentum. One of the difficulties in
trying to study this using GWD is the problem of the hard boundary. A hard
boundary can be simulated using a polynomial bounding potential of a high degree
such as $V(x) = (x/L)^{2d}$ where $2L$ is the length of the box and $d$ is a
sufficiently large number. GWD can only approximately account for the penetration
of the quantum wave packet into classically forbidden regions if the potential
is anharmonic. Therefore, the presence of the hard wall makes the simulations
more inaccurate. Ideas from the general Gaussian wavepacket dynamics (GGWD)
method~\cite{huberGeneralizedGaussianWave1987, huberGeneralizedGaussianWave1988}
will be leveraged in the future to make the simulations more accurate.

\section{Conclusion}\label{sec:conclusion}
We have used Heller's thawed Gaussian wavepacket
dynamics~\cite{hellerTimeDependentApproach1975,
    hellerTimeDependentVariational1976} to analyse the dynamics of non-Hermitian
continuum quantum systems. While recent numerical studies have described curious
features of the dynamics~\cite{liDynamicSkinEffects2022}, a semiclassical
analysis provides a general intuitive framework for thinking about these
systems, and makes the connections to classical mechanics more obvious. As a
problem of interest, we have chosen a generalized continuum Hatano-Nelson model
in an external potential, with a potentially position-dependent non-Hermitian
term. The non-Hermitian term appears as an imaginary vector potential.

We show that in general, due to the presence of the imaginary vector potential,
the dynamics requires the use of complex phase space for specifying the
time-evolving Gaussian wavepacket. However, using the non-uniqueness of these
generalized Gaussian wavepackets, one can transform the initial condition in
such a manner that the entire classical trajectory has real positions, obtained
from the standard Newtonian equations of motion involving only the external
potential. This ``guiding'' classical trajectory however has complex momenta. We
show that the Newtonian equation of motion is independent of the vector
potential, which affects the dynamics in three ways. The first way, which
affects the guiding trajectory, is that the transformed initial condition is
dependent on the imaginary vector potential. Second, the equations of motion for
$\alpha(t)$ and $\gamma(t)$ are dependent on the vector potential. The third way
by which the non-Hermitian term affects the dynamics is that on transforming
back to fully real phase space coordinates, we notice that the center of the
wavepacket deviates from the guiding trajectory by an amount that is
proportional to the strength of the vector field at that point. The
proportionality constant is the instantaneous width of Gaussian wavepacket.
Thus, we have shown that the sharper the wavepacket and the weaker the
non-Hermitian term, the closer the center is to the position of the guiding
trajectory. It is interesting that the non-Hermitian term enters the dynamics by
modifying $\alpha(t)$ and also through the instantaneous width of the
wavepacket.

The interaction between the external potential and the non-Hermitian term
becomes manifestly clear through this decomposition of the real phase space
trajectory in terms of the guiding trajectory. We show that for a constant
vector potential, the analysis becomes even more transparent, since $\alpha(t)$
becomes only dependent on the external potential. Then the spread remains
exactly the same as that of the corresponding Hermitian system. The center of
the wavepacket deviates by the width of the wavepacket times the non-Hermitian
term.

The GWD approach provides a novel way of conceptualizing the dynamics under
non-Hermitian systems. The decomposition in terms of the guiding classical
trajectory and the deviations therefrom gives a classical picture to the
dynamics. The GWD described in this work also provides the basis for future work
that uses generalized Gaussian wavepackets for improved
propagation~\cite{huberGeneralizedGaussianWave1987,
    huberGeneralizedGaussianWave1988}. Additionally, trying to understand the
dynamics of Gaussian wave packets in paradigmatic examples of
$\mathcal{PT}$-symmetric non-Hermitian Hamiltonians would further develop our
fundamental understanding of these interesting systems.

\begin{acknowledgments}
    I would like to thank Awadhesh Narayan for introducing me to non-Hermitian
    systems and for multiple discussions.
\end{acknowledgments}

\bibliography{library}
\end{document}